\documentclass[journal]{vgtc}                
\ifpdf
  \pdfoutput=1\relax                   
  \usepackage{graphicx}                
  \DeclareGraphicsExtensions{.pdf,.png,.jpg,.jpeg} 
\else
  \usepackage{graphicx}                
  \DeclareGraphicsExtensions{.eps}     
\fi%

\graphicspath{{figures/}{pictures/}{images/}{./}} 

\usepackage{microtype}                 
\usepackage{hyperref}
\PassOptionsToPackage{warn}{textcomp}  
\usepackage{textcomp}                  
\usepackage{mathptmx}                  
\usepackage{times}                     
\usepackage{cite}                      
\usepackage{tabu}                      
\usepackage{booktabs}                  

\onlineid{0}

\vgtccategory{Research}
\vgtcpapertype{system}
\newcommand{\thename}{VisMCA}
\title{\thename{}: A Visual Analytics System for Misclassification\\ Correction and Analysis

\LARGE{VAST Challenge 2020, Mini-Challenge 2 Award: Honorable Mention for Detailed Analysis of Patterns of Misclassification}}

\author{Huyen N. Nguyen, Jake Gonzalez, Jian Guo, Ngan V.T. Nguyen, and Tommy Dang}
\authorfooter{
\item
 The authors are with the Department
of Computer Science, Texas Tech University, Lubbock,
TX, USA. E-mail: \{huyen.nguyen\,$|$\,jake.gonzalez\,$|$\, jian.guo\,$|$\,ngan.v.t.nguyen\,$|$\,tommy.dang\}@ttu.edu.
\item Huyen N. Nguyen and Jake Gonzalez contributed equally to this work.
}

\shortauthortitle{Biv \MakeLowercase{\textit{et al.}}: Global Illumination for Fun and Profit}

\abstract{This paper presents \thename{}, an interactive visual analytics system that supports deepening understanding in ML results, augmenting users' capabilities in correcting misclassification, and providing an analysis of underlying patterns, in response to the VAST Challenge 2020 Mini-Challenge 2. 
\thename{} facilitates tracking provenance and provides a comprehensive view of object detection results, easing re-labeling, and producing reliable, corrected data for future training. Our solution implements multiple analytical views on visual analysis to offer a deep insight for underlying pattern discovery.} 


\CCScatlist{ 
 \CCScat{K.6.1}{Management of Computing and Information Systems}%
{Project and People Management}{Life Cycle};
 \CCScat{K.7.m}{The Computing Profession}{Miscellaneous}{Ethics}
}

\teaser{
  \centering
  \includegraphics[width=0.95\linewidth]{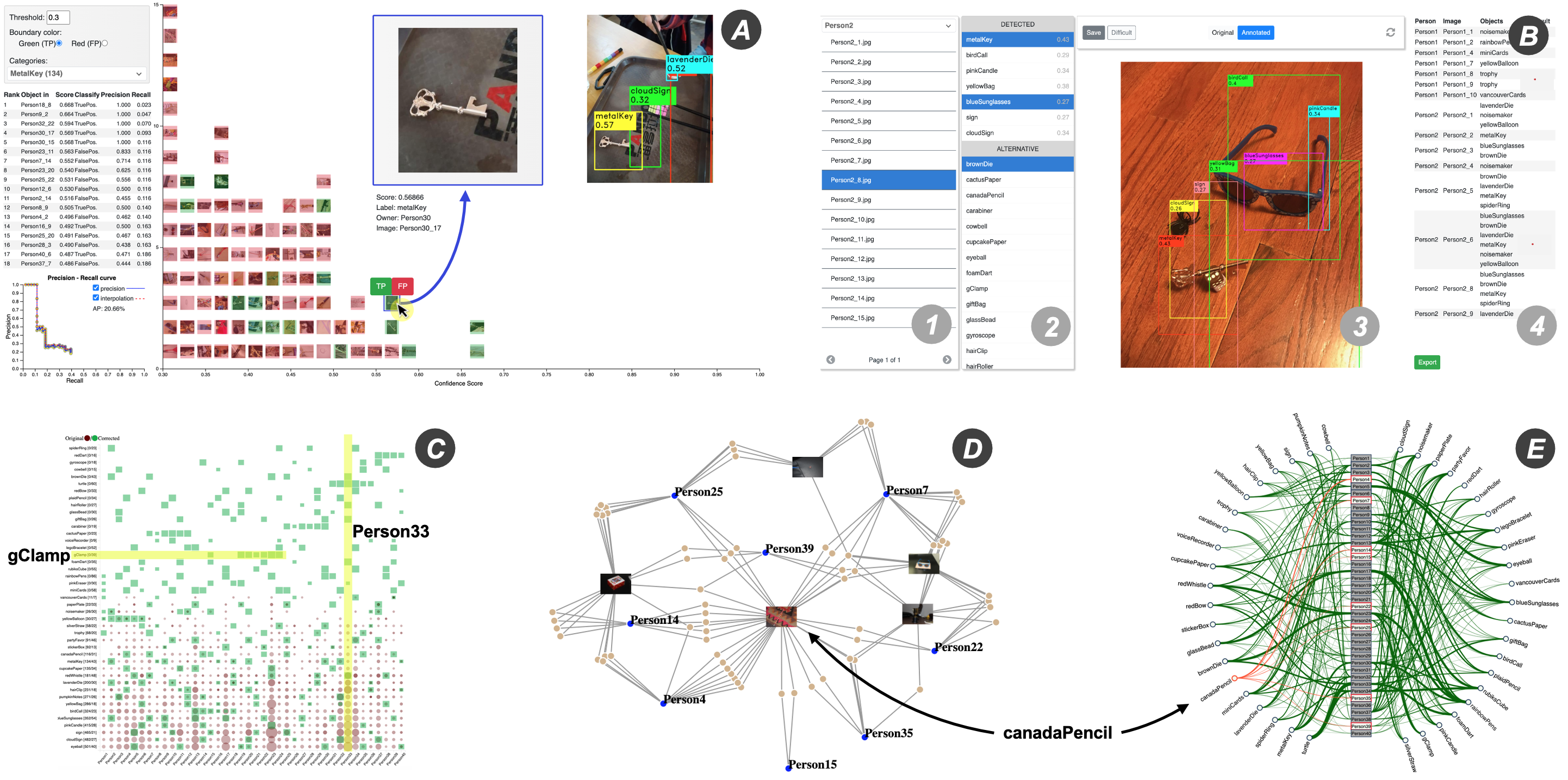}
  \caption{The \thename{} interface. The system presents the distribution of detections and allows real-time correction by the user (a). The image correction pipeline (b) supports label annotation and resulting data export. The matrix (c) visualizes the original and corrected labels with regards to respective owners. 
  The filtered network of people sharing the same objects revolves around \textit{canadaPencil} (d). The list-with-links repurposed network (e) provides an interactive visualization to identify the two-way owner-object relationships.}
  \label{fig:teaser}
}
\begin{document}



\maketitle


\section{Introduction}
\label{sec:intro}

The VAST Challenge 2020 Mini-Challenge 2 asked participants to uncover the relationships between people, objects, and image data retrieved from a social media platform, ultimately leading to identify the ``totem''--a secret item shared within an unknown subgroup. There are 900 images, belonging to 40 people and describing 43 classes, running through a machine learning (ML) model to identify objects. Visualization techniques have previously been applied towards the explainability of the ML model~\cite{deepvix, le2019visualization}. In this paper, we utilize visualization and visual analytics to correct and analyze misclassification.

We developed \thename{}, a comprehensive web-based visual analytics system that facilitates human-in-the-loop in misclassification correction and analysis, implemented with Python, HTML5, and D3js~\cite{d3}. The systematic interface of \thename{} is shown in Figure \ref{fig:teaser}. The general workflow includes 1) characterize the detection distribution across different levels of confidence score, then 2) correct misclassification and produce data for future ML uses, and 3) analyze underlying patterns via multiple analytical views. The complete report and demonstration are available online at  \url{https://huyen-nguyen.github.io/VAST2020mc2/}.





\section{System Architecture}

\subsection{Detection Distribution}
The overall view is presented in Figure \ref{fig:teaser}A. The detection distribution provides a quick overview of images with regards to different confident score threshold. Regarding interaction, users can quickly cut to images with low initial confidence, mark true positives and false positives and update results on-the-fly by a simple right-click, and bulk select within a responsive layout. We use Average Precision - precision averaged across all recall values to evaluate each classifier's performance.

\subsection{Misclassification Correction}
Figure \ref{fig:teaser}B describes the pipeline for correcting classification errors. Panel 1 consists of image and person selection dropdown lists. Panel 2 contains a list for selecting and deselecting labels, categorized into: Detected--labels from the image's prediction, sorted by confidence score, and Alternative--all the other labels, sorted alphabetically. Sorting aid users to quickly browse and select relevant items~\cite{Shneiderman1996TheEH}. Panel 3 displays the image frame, which includes two modes for displaying images: Original and Annotated with labels, bounding boxes, and confidence score, providing users with explicit illustration of the detections. Users can click the ``Save'' button to save the assigned labels, whose effect is updated on-the-fly on the data table in Panel 4.

With the ``Export'' feature, the re-labeled results can be exported in CSV format for further use. If the user is unsure about their manual annotation on some example, they can mark as \textit{``Difficult.''} This feature provides value to ML algorithms and future users of the results, as \textit{difficult} examples can be treated differently in different training settings.

\subsection{Object Distribution Matrix}
The object distribution matrix shows how objects are distributed amongst the 40 suspected group members, as shown in Figure \ref{fig:teaser}C. The original data is depicted by brown circles whose size depends on the number of detections, and opacity is determined by the average of the confidence scores across all images containing that object. The corrected data is depicted by the green squares whose size represents an object's occurrences in a person's images. The people and objects are presented horizontally and vertically, along with statistics of the occurrences in images in the original and corrected dataset.

\subsection{Network of Top Object}
To gain a clear view of the interconnection between different entities, we built a complete network of people and their objects. For demonstration purpose, we present the filtered network revolving around \textit{canadaPencil}, as shown in Figure \ref{fig:teaser}D. Filtering is useful to focus on the points of interest by eliminating unwanted items~\cite{Shneiderman1996TheEH}, especially in a complex network~\cite{van2019hackernets}, where we need to detect communities~\cite{eqsa}. The blue and yellow nodes represent people and objects, respectively, where nodes representing the same object are linked to a reference image.

\subsection{List-with-links Repurposed Network}
The visualization is shown in Figure \ref{fig:teaser}E, in which people are shown in a vertical column, and objects are arrayed in a circle. The curved green links use width to indicate the number of detections: the number of images containing an object belongs to a person, highlighted in orange on hovering. This visualization is useful to quickly find groups of people that shared objects and the set of objects that one owns.

\section{Results}

\begin{figure}
 \centering 
 \includegraphics[width=\columnwidth]{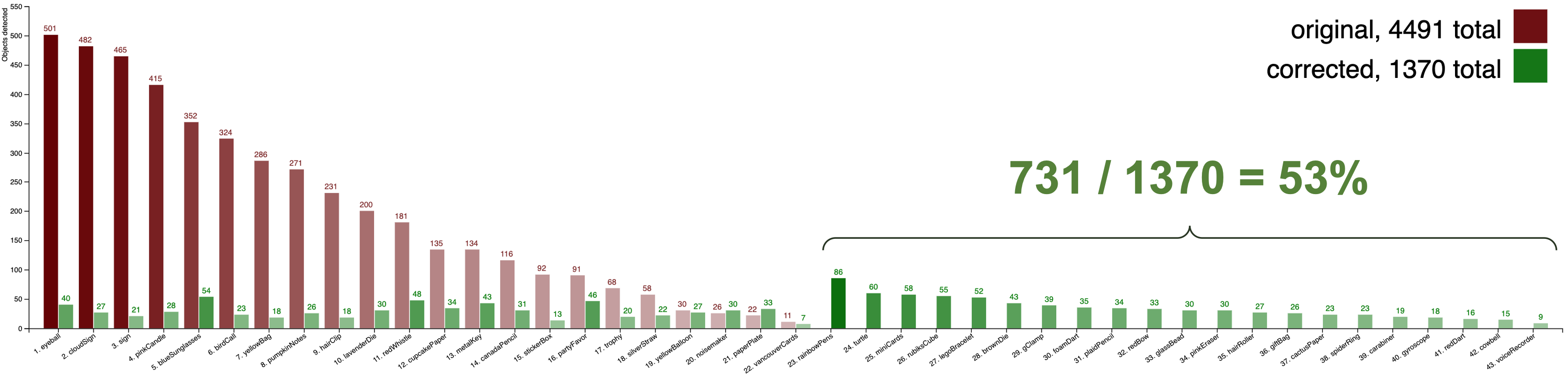}
 \caption{Distribution of classes before and after correction process.}
 
 \label{fig:ov}
\end{figure}

There are a total of 43 classes in the training set, distribution quite evenly among the 900 images. However, there are detections in only 22 classes and no detection for any of the remaining classes. This can be seen from the matrix in Figure \ref{fig:teaser}C, where the upper half of the matrix shows no detection (no brown circles), while there actually are objects in those images (green squares). The ML model missed out at least 53\% of the objects, as shown in Figure~\ref{fig:ov}, hence not a good model. As highlighted in Figure \ref{fig:teaser}C, from \textit{Person19} to \textit{Person24} having \textit{gClamp}. For \textit{Person33}, only one object was correctly detected (one overlap); all other detections in this person's images are false positive.

\begin{figure}[tb]
 \centering 
 \includegraphics[width=\columnwidth]{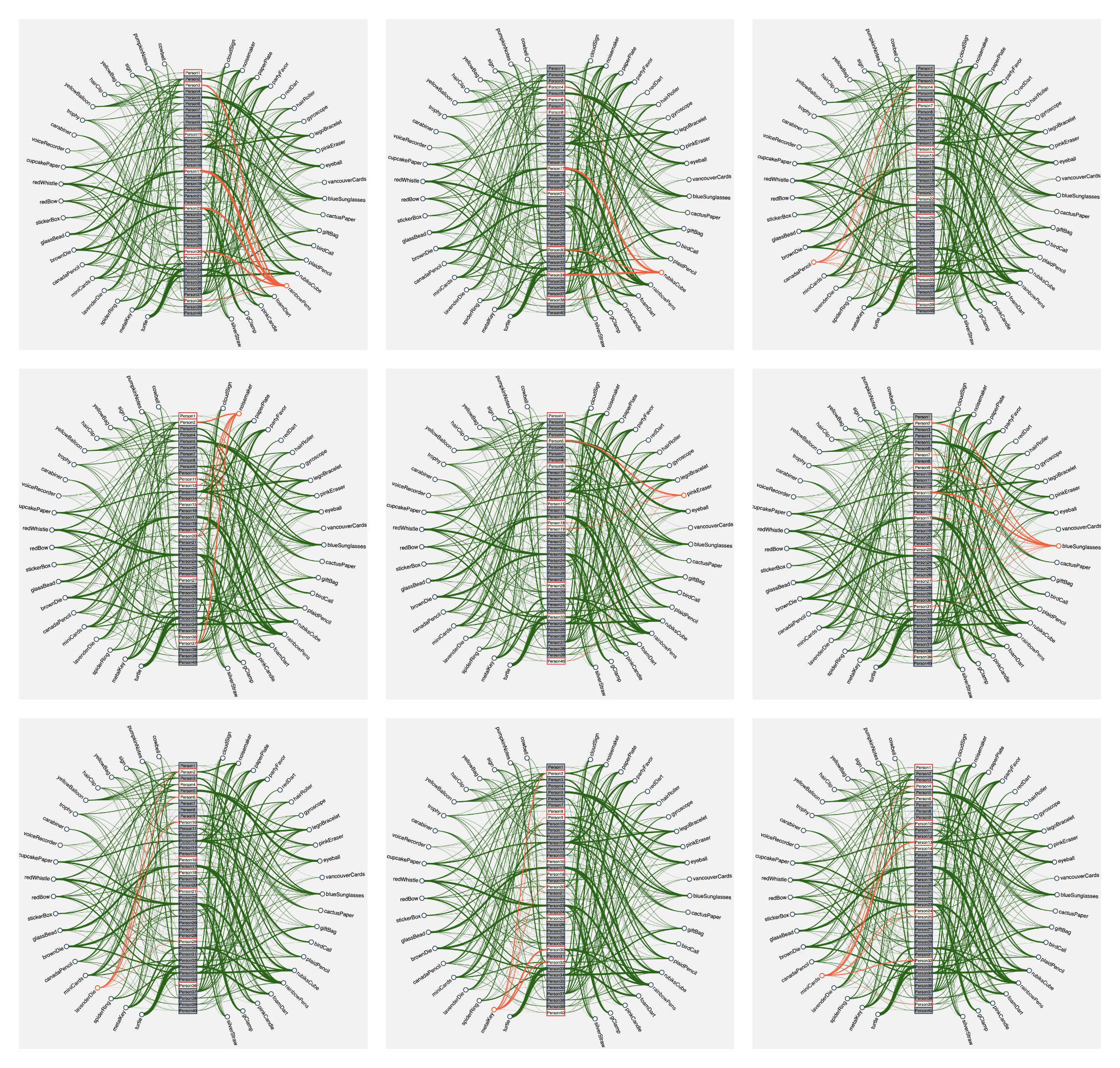}
 \caption{The networks of 9 objects that each shared by \textit{at least} 8 people.
 The objects are: \textit{rainbowPens, rubiksCube, canadaPencil, noisemaker, pinkEraser, blueSunglasses, lavenderDie, metalKey, miniCards.}}
 
 \label{fig:nine}
\end{figure}

As stated in the challenge description, there are eight people connected to the totem. As shown in Figure~\ref{fig:nine}, there are four objects owned by only eight people: \textit{canadaPencil}, \textit{rainbowPens}, \textit{noisemaker}, and \textit{rubiksCube}. We speculated that each person should have at least two images of the totem, representing sending and receiving signals. Among those four objects, only \textit{canadaPencil} owners have two or more images per person, while the rest includes owners having one image, which concludes that \textit{canadaPencil} is the totem.

\section{Conclusion and Future work}
This paper presents \thename{}, an interactive visual analytics system for a comprehensive view of ML results: augmenting users' capabilities in correcting misclassification and providing an analysis of underlying patterns. To make the correction process more efficient, we suggested considering overlapping bounding boxes and common object combinations. This approach might require validation testing and verification in future work, along with addressing uncertainty in classification results.


\bibliographystyle{abbrv-doi}

\bibliography{template}
\end{document}